\documentclass [12pt]{article}
\usepackage{amsmath,amsfonts}
\newcommand{\IR}{\mathbb{R}}

\newcommand{\IC}{\mathbb{C}}

\begin{document}
\begin{titlepage}
\begin{center}
{\Large\bf Semiclassical Limit of the Dirac Equation\vspace{5mm}\\
and Spin Precession\vspace{2cm}}\\
{\large Herbert Spohn} \medskip\\
Zentrum Mathematik and Physik Department, \\
TU M\"unchen, D--80290 M\"{u}nchen, Germany
 {\tt spohn@mathematik.tu-muenchen.de}
\end{center} \vspace{6cm}
{\bf Abstract}. We study the Dirac equation with slowly varying external potentials.
Using matrix-valued Wigner functions we prove that the electron follows
with high precision the classical orbit and that the spin precesses according to
the BMT equation with gyromagnetic ratio $g=2$.
\end{titlepage}

\section{Introduction} \label{a}

The high precision experiments of the anomalous magnetic moment of
the electron or the muon use in their interpretation the BMT
equation for spin precession \cite{bai}. The derivation of
Bargmann, Michel, and Telegdi \cite{barg} does not even mention
quantum mechanics, however. Rather their arguments are based on
relativistic covariance together with a classical notion of an
intrinsic angular momentum. Thus one might wonder how exactly the
classical BMT equation is connected with the quantum mechanical
spin precession. Properly speaking, one should start from QED and
then deduce a BMT-like equation for the spin motion. Such a
program looks rather difficult at present. But even if we take the
one-particle relativistic Dirac equation as starting point, as we
do here, the problem seems little explored and is thus still of
interest.

Let us first recall the BMT equation. One considers a charge, mass
$m$, charge $e$, moving in prescribed, time-independent external
electromagnetic fields $\boldsymbol{E}= -\nabla \phi$, $\boldsymbol{B} =
\nabla \times \boldsymbol{A}$
and governed by the Lorentz force equation
\bigskip
\begin{equation}\label{a.a}
\frac{d}{dt}m\gamma \boldsymbol{v}_{t}= e \big(\boldsymbol{E}(\boldsymbol{q}
_{t})+ c^{-1} \boldsymbol{v}_{t}\times \boldsymbol{B}(\boldsymbol{q}_{t})\big)\,.
\end{equation}
Here $\boldsymbol{q}_{t}$ is the position and $\boldsymbol {v}_{t}= 
\dot{\boldsymbol{q}}_{t}$
the velocity of the
particle at time $t$ with $\gamma = 1/\sqrt{1-(\boldsymbol{v}/c)^2}$. 
According to
BMT, for $g=2$, the precession of the spin $\boldsymbol{s}$ along the 
given orbit is determined by
\begin{equation}\label{a.b}
\frac{d}{dt} \boldsymbol{s} = \frac{e}{mc}\,\boldsymbol{s} \times
\left[\frac{1}{\gamma}\boldsymbol{B}-\frac{1}{1+\gamma} c^{-1} \boldsymbol{v}
\times \boldsymbol{E}\right].
\end{equation}
For $g \not= 2$ the prefactors of $\boldsymbol{B}$, $\boldsymbol{v} \times
\boldsymbol{E}$ change and also a term
proportional to $\boldsymbol{v}(\boldsymbol{v} \cdot \boldsymbol{B})$ appears.
We refer to the textbook by Jackson,  Section 11.11, for details \cite{jack}.

The same physical situation can be described by
the relativistic one-particle Dirac equation in the semiclassical
limit. This means that the potentials $\phi, \boldsymbol{A}$ are slowly varying
on the scale set by the typical extension of the electron/positron
wave function. On a formal level it is convenient to
introduce the dimensionless scale parameter $\varepsilon $, $ \varepsilon
\ll 1$, and to take the potentials to be of the form
\begin{equation}\label{a.c}
\phi(\varepsilon \boldsymbol{x})\, ,\;\; \boldsymbol{A}(\varepsilon 
\boldsymbol{x})\,.
\end{equation}
The state of the electron/positron is now described by the four
component spinor $\psi$ which evolves according to the Dirac equation
as
\begin{equation}\label{a.d}
i \hbar \frac{\partial}{\partial t}\psi = H \psi
\end{equation}
with the Dirac hamiltonian
\begin{equation}\label{a.e}
H = c\gamma_0 \big(\boldsymbol{\gamma} \cdot (-i\hbar \nabla_x - \frac{e}{c} \boldsymbol{A}
(\varepsilon \boldsymbol{x})) + mc \big) + e \phi (\varepsilon \boldsymbol{x})
\end{equation}
acting as a linear operator on $\bigoplus^4_{n=1} L ^2(\IR^3)$. 
Here  $\gamma _0 ,\boldsymbol\gamma$ are the Dirac gamma
matrices, where we follow the conventions in \cite{yndu}. For the
semiclassical limit one starts with a well localized wave function
$\psi$, having zero positron component, and evolves it unitarily
according to
\begin{equation}\label{a.f}
\psi_{t} = e^{-iHt/\hbar}\psi
\end{equation} over such a long time
that the response to the external forces is visible. With our
choice (\ref{a.c}) this means times of the order $\varepsilon^{-1}$. If the
standard picture is correct, then over that time-span $\psi_{t}$ should
have a negligible positron component and should be well
concentrated on the classical path determined by (\ref{a.a}). In
addition, the average polarization should satisfy (\ref{a.b}). This is
exactly what we are going to show.

Before entering in the details of our argument, it might be useful
to recall the semiclassical limit of the Schr\"{o}dinger equation
\begin{equation}\label{a.g}
i \hbar \frac{\partial}{\partial t} \psi = \big(-\frac{\hbar^{2}}{2m}
\Delta + V
(\varepsilon \boldsymbol{x})\big)\psi
\end{equation}
with the slowly varying electrostatic potential $V$. If the initial
$\psi$ is well localized at the classical phase point $(\boldsymbol{q},
\boldsymbol{p})$,
then $\psi_{t}$ is well localized at $(\boldsymbol{q}_{t}, 
\boldsymbol{p}_{t})$ with
$\dot{\boldsymbol{q}}= \frac{1}{m}\boldsymbol{p}$, 
$ \dot{\boldsymbol{p}} = -\nabla V
(\boldsymbol{q})$.
This is an extremely well-studied chapter of mathematical physics 
\cite{rob} with higher order
corrections and sharp error bounds available \cite{hag1}. 
The Schr\"{o}dinger equation is scalar.
Put it differently, if we set the potential to zero, then the
energy function of (1.7) is $E(\boldsymbol{p}) = \frac{1}{2m} 
\boldsymbol{p}^2$ which is a
scalar function. In somewhat more complicated cases, like an
additional short scale periodic potential \cite{hoev}, internal
degrees of freedom, or several particles, for zero external
potential the energy will be matrix-valued as a function of a
suitable quasimomentum.
If this energy matrix has nondegenerate eigenvalues (= energy band
functions), then in essence one is back to the scalar case.
Interference between bands is destroyed through rapid oscillations
and the evolution in each band separately is semiclassical
\cite{hoev}. In general, energy bands may cross, i.e. the
eigenvalues may be degenerate along lower dimensional
submanifolds. The wave function can then propagate into
neighboring bands giving rise to intricate interference patterns
\cite{hage}. In the case of the Dirac equation, the energy matrix
is a $4 \times 4$ matrix and it has the two energy bands $E_\pm
(\boldsymbol{p}) = \pm c \sqrt{m^2c^2 + \boldsymbol{p}^2}
$, both two-fold degenerate, which is at the origin for a scenario
very distinct from the scalar case. In fact, in the semiclassical
limit, the position and momentum become classical variables,
whereas the spin remains fully quantum mechanical.

It should be noted that from a purely abstract point of view the
Dirac equation is the simplest example for a matrix-valued energy
function. One would like to have at least two bands each of which
should be degenerate. This makes four to be the minimal matrix
dimension.

We close the introduction with some brief comments on previous
work. The most elementary approach is to consider average values
of spin and position and to use what is known as Ehrenfest's
theorem. For the Dirac equation the details have been carried
through by Fradkin and Good \cite{frad}. The more sophisticated
WKB approach to the Dirac equation was initiated by Pauli
\cite{paul} and completed by Rubinow and Keller \cite{rubi}. The
full semiclassical, van-Vleck-type propagator has been worked out
recently by Bolte and Keppeler \cite{bolte} including a trace
formula for the eigenvalues. The WKB method has the drawback of
being local in time only. In the case of the Schr\"{o}dinger
equation one understands how to continue beyond the caustics by
adapting new coordinates in the so-called Lagrangian manifold
\cite{rob,masl}. For matrix-valued energy functions such a program 
has not
been attempted yet. Instead, we will use here  matrix-valued Wigner 
functions. The abstract theory for this approach has been developed in
\cite{gera} and applications to beam physics are discussed in
\cite{heine}.

\section{Semiclassical limit of the Dirac equation}\label{b}
\setcounter{equation}{0}
In the Appendix we study abstractly the semiclassical limit for
Schr\"{o}dinger type equations with matrix-valued symbols. In this
section we merely transcribe these general results to the Dirac
equation. The emergence of the BMT equation will be explained in
Section \ref{c}.

In the following $\boldsymbol{p} $ denotes the classical momentum and we
use $\boldsymbol{q}, \boldsymbol{p}$ as a pair of canonical coordinates. 
$\{\cdot , \cdot\}$ is the Poisson
bracket. We define it also for matrix-valued functions $A, B$
through
\begin{equation}\label{b.a}
\{A, B\} = \nabla_p A \cdot \nabla_q B -
\nabla_q A \cdot \nabla_p B
\end{equation}
which is again a matrix-valued function. Here $\cdot$ refers to the
scalar product between the gradients. Note that the order of
factors must be respected. In particular,
\begin{equation}\label{b.b}
\{ A, A \} \not= 0
\end{equation}
in general.

The Dirac matrix reads
\begin{equation}\label{b.c}
H_D = c \gamma_0 \big(\boldsymbol{\gamma} \cdot (\boldsymbol{p} - \frac{e}{c}
\boldsymbol{A}(\boldsymbol{q})) + mc \big) + e \phi
(\boldsymbol{q})\, ,
\end{equation}
compare with (\ref{a.e}), which is regarded as a $4 \times 4$ matrix-valued
function on the classical phase space, $(\boldsymbol{q},\boldsymbol{p}) 
\in \IR^3
\times \IR^3$. $H_D$ is easy to diagonalize. The eigenvalues are
\begin{equation}\label{b.d}
h_\pm = \pm c p_0 + e \phi
\end{equation}
and the corresponding two-dimensional eigenprojections
\begin{equation}\label{b.e}
P_\pm = \frac{1}{2}\, \big(1 \pm \frac{1}{p_0} 
\gamma_0 (\boldsymbol{\gamma} \cdot
(\boldsymbol{p} - \frac{e}{c} \boldsymbol{A}) + mc)\big)\, ,
\end{equation}
where $p_0 = \sqrt{m^2 c^2 + (\boldsymbol{p} - \frac{e}{c} \boldsymbol{A})^2}$. 
Thus
\begin{equation}\label{b.f}
H_D = h_+ P_+ + h\_P\_ \,.
\end{equation}
$P_+$ is the electron and $P\_$ the positron subspace.
In addition, we need the polarization matrix along the orientation
$\boldsymbol{a},\boldsymbol{ | a |} = 1$,
which is given by
\begin{equation}\label{b.g}
\boldsymbol{a} \cdot \boldsymbol{S} = (\boldsymbol{\gamma} \cdot \boldsymbol{g}
+ \frac{1}{p_0} \boldsymbol{a} \cdot (\boldsymbol{p} -
\frac{e}{c} \boldsymbol{A})) \gamma_5
\end{equation}
with
\begin{equation}\label{b.h}
\boldsymbol{g} = \boldsymbol{a} - \frac{1}{p_0 (p_0 + mc)}
(\boldsymbol{a} \cdot(\boldsymbol{p} - \frac{e}{c}\boldsymbol{A}))(\boldsymbol{p} -
\frac{e}{c}\boldsymbol{A})\,,
\end{equation}
cf. \cite{frad}. $\boldsymbol{a} \cdot \boldsymbol{S}$ has eigenvalues $\pm 1$,
$(\boldsymbol{a} \cdot \boldsymbol{S})^2 = 1$, and commutes with
$H_D$,
\begin{equation}\label{b.i}
[H_D , \boldsymbol{S}] = 0 \,.
\end{equation}

We discuss only the motion of an electron, the positron subspace
being simply its mirror image. In the semiclassical limit the state of 
the electron in general
is described through the matrix-valued Wigner function 
$W_{t}(\boldsymbol{q}, \boldsymbol{p})$
on classical phase space. According to the Appendix, 
$W_{t}$ is governed by 
\begin{equation}\label{b.l}
\frac{\partial}{\partial t} W_{t} (\boldsymbol{q}, \boldsymbol{p}) 
= - \{h_+
(\boldsymbol{q}, \boldsymbol{p}), P_+ W_{t}
(\boldsymbol{q}, \boldsymbol{p})\} - i [H^{(+)}_s (\boldsymbol{q}, 
\boldsymbol{p}),
W_{t} (\boldsymbol{q}, \boldsymbol{p})]
\end{equation}
with
\begin{eqnarray}\label{b.m}
H^{(+)}_s  &=&  -i \big( [P_+, \{h_+, P_+\}]  \nonumber \\ 
&&+\frac{1}{2} P_+ (h_+ \{P_+,
P_+\} -  h_- \{P_-\, , P_-\}) P_+ \big) = H_{be} + H_{nn}\,.
\end{eqnarray}
If the initial wave packet is well localized at $\boldsymbol{q}^{0},
\boldsymbol{p}^{0}$,
then one has to solve (\ref{b.l}) with the initial condition 
$W_{0}(\boldsymbol{q},
\boldsymbol{p})= \delta (\boldsymbol{q}-\boldsymbol{q}^{0}) \delta 
(\boldsymbol{p}-
\boldsymbol{p}^{0}) |\varphi_{0}\rangle\langle \varphi_{0}|$ and 
$\varphi_{0} \in  P_{+} \IC^4$.
In the course of time, this structure is preserved,
\begin{equation}\label{b.n}
    W_{t}(\boldsymbol{q},
\boldsymbol{p})= \delta (\boldsymbol{q}-\boldsymbol{q}_{-t}) \delta 
(\boldsymbol{p}-
\boldsymbol{p}_{-t}) |\varphi_{t}\rangle\langle\varphi_{t}|\, ,
\end{equation}
where $\boldsymbol{q}_{t},\boldsymbol{p}_{t}$ is the solution to 
\begin{equation}\label{b.j}
\dot{\boldsymbol {q}} = \{h_+ ,\boldsymbol q\} = \nabla_p h_+ \,,  
\quad \dot{\boldsymbol{p}}
= \{h_+ ,\boldsymbol{p}\} = - \nabla_q h_+ 
\end{equation}
with initial conditions $\boldsymbol{q}^{0},\boldsymbol{p}^{0}$. 
(\ref{b.j}) is equivalent to (\ref{a.a}) with $p_0 = mc\gamma$ and the velocity
defined through
\begin{equation}\label{b.k}
\boldsymbol{v} = \frac{1}{p_0} \,( c \boldsymbol{p} - e 
\boldsymbol{A})\, .
\end{equation}
The spinor $\varphi_{t}$ evolves with a time-dependent hamiltonian 
according to
\begin{equation}\label{b.oo}
 i \frac{d}{dt} \varphi_t = H_{s}^{+}(\boldsymbol{q}_{t},\boldsymbol{p}_{t})
 \varphi_t \, .
\end{equation}    
By (\ref{b.m}) this implies $\varphi_{t} \in  P_{+}(\boldsymbol{q}_{t},
\boldsymbol{p}_{t}) \IC^4$ for all $t$.

In (\ref{b.l})
the Poisson bracket $\{h_+, \cdot \}$ transports the spinor along the
classical orbit maintaining its
orientation. Therefore, in general, the spinor would move out of the
electron subspace which is precisely compensated for by $H_{be}.$
For that reason, the analogue of $H_{be}$ has been baptized Berry
term in \cite{little}, while the remainder, $H_{nn}$, has ``no name'' and
is referred to also as Poissonian curvature in \cite{emm}.

It remains to compute $H^{(+)}_s$. $H_{be}$ is actually not needed in explicit form, but
we list it for completeness,
\begin{equation}\label{b.o}
H_{be}=\frac{e}{2 p_0} (\gamma_5  \gamma_0  \boldsymbol {\gamma}  \cdot (c^{-1}
\boldsymbol{v} \times \boldsymbol{F}) - i \frac{1}{p_0} \boldsymbol \gamma \cdot \boldsymbol{F})
\end{equation}
with $\boldsymbol{F} = \boldsymbol{E} + c^{-1} \boldsymbol{v} 
\times \boldsymbol{B}$.
For the no name term, we note that
\begin{equation}\label{b.p}
\{P_+, P_+\} = \{1-P\_\,, 1-P\_\} = -i \frac{e}{2cp^2_0} \gamma_5 
\gamma_0 \boldsymbol{\gamma} \cdot \boldsymbol{B}
\end{equation}
and hence
\begin{eqnarray}\label{b.q}
H_{nn} &=& \frac{1}{2} i P_+ \big((-cp_0 - e \phi) \{P_+, P_+\} + (-cp_0 +
e \phi) \{P\_\,, P\_\}\big) P_+ \,\nonumber \\ &=& - \frac{e}{2p_0} P_+
\gamma_5  \gamma_0  \boldsymbol{\gamma}  \cdot \boldsymbol{B} P_+ \,.
\end{eqnarray}

\section{Spin precession}\label{c}
\setcounter{equation}{0}
In the semiclassical limit the average electron polarization
along $\boldsymbol{a}$ is given by
\begin{equation}\label{c.a}
\boldsymbol{a} \cdot \boldsymbol{s}_{t} = \int d^3 q d^3 p\, \mathrm{tr}
[W_{t} (\boldsymbol{q}, \boldsymbol{p}) P_+ (\boldsymbol{q}, \boldsymbol{p})
\boldsymbol{a}\cdot \boldsymbol{S}(\boldsymbol{q}, \boldsymbol{p})
P_+(\boldsymbol{q}, \boldsymbol{p})]
\end{equation}
with $W_{t}$ the matrix-valued Wigner function. Note 
that $[P_+, \boldsymbol{S}] =
0$. According to the Appendix, we have
\begin{eqnarray}\label{c.b}
\frac{d}{dt}  \boldsymbol{a} \cdot \boldsymbol{s}_{t} & = & \frac{d}{dt} \int d^3 q 
d^3 p\, 
{\rm tr} [W_{t} P_+ \boldsymbol{a} \cdot \boldsymbol{S}] \\ & = & \frac{1}{2}\int d^3 q\,
d^3 p \,{\rm tr} [P_+ ( \{W_{t}, H_D\} - \{H_D, W_{t}\}) P_+ \boldsymbol{a} 
\cdot
\boldsymbol{S}] \nonumber \\ & = &
\int d^3 q \, d^3 p \,{\rm tr}[P_+ (\{W_{t}, h_+\} + i 
[W_{t}, H_{nn}]) P_+ \boldsymbol{a} \cdot
\boldsymbol{S}]
 \nonumber \\ & = & \int d^3 q \, d^3 p \, {\rm tr} [W_{t} 
 (\{h_+, P_+ \boldsymbol{a} \cdot
\boldsymbol{S} P_+\} + i [H_{nn}, P_+ \boldsymbol{a} \cdot \boldsymbol{S}
P_+])]\, .\nonumber
\end{eqnarray}

Using $P_+ \{h_+ , P_+\} P_+ = 0$ we have
\begin{equation}\label{c.e}
\frac{d}{dt} \boldsymbol{a} \cdot \boldsymbol{s}_{t} = \int d^3 q  d^3 p 
\, {\rm tr}[W P_+
(\{h_+, \boldsymbol{a} \cdot \boldsymbol{S}\} - i \frac {e}{2 p_0} [\gamma_5 
\gamma_0
\boldsymbol{\gamma} \cdot \boldsymbol{B}, \boldsymbol{a} 
\cdot \boldsymbol{S}] ) P_+]
\end{equation}
and only have to compute the term $P_+ (\dots) P_+$, which is
accomplished using the identities
\begin{equation}\label{c.f}
(p_0 / mc) P_+ \boldsymbol{v} \cdot \boldsymbol{\gamma}  \gamma_5  P_+ = P_+ \boldsymbol{v}
\cdot \boldsymbol{S} P_+ = c P_+  \gamma_5 P_+ \, .
\end{equation}
For the Berry term we have
\begin{eqnarray}\label{c.g}
\lefteqn{P_+ \{h_+ , \boldsymbol{a} \cdot \boldsymbol{S}\} P_+ = P_+
\boldsymbol{a} \cdot \dot{\boldsymbol{S}} P_+ = P_+
((\boldsymbol{\gamma} \cdot \dot {\boldsymbol{g}} + c^{-1} \boldsymbol{a} \cdot
\dot{\boldsymbol{v}}) \gamma_5) P_+}
\nonumber \\ &=& c^{-2} \big( - \frac{p_0^2}{(mc + p)^2}  c^{-2} (\boldsymbol{a}
\cdot \boldsymbol{v}) (\boldsymbol{v} \cdot \dot{\boldsymbol{v}})
+ \frac{p_0}{mc + p_0} (\boldsymbol{a} \cdot \dot{\boldsymbol{v}})\big) P_+ \boldsymbol{v}
\cdot \boldsymbol{S} P_+ \nonumber \\&& - c^{-2} \frac{p_0}{mc + p_0}(\boldsymbol{a}
\cdot \boldsymbol{v}) P_+ \boldsymbol{\gamma} \cdot \dot{\boldsymbol{v}}
P_+\, ,
\end{eqnarray}
where we used (\ref{c.f}). The last term transforms as
\begin{equation}\label{c.h}
P_+ \boldsymbol{\gamma} \cdot \dot{\boldsymbol{v}} P_+ = P_+\dot{\boldsymbol {v}}
\cdot \boldsymbol{S}  P_+ -\frac{p_0}{mc + p_0} c^{-2} (\boldsymbol{v}
\cdot \dot{\boldsymbol{v}}) P_+ \boldsymbol{v} \cdot \boldsymbol{S} P_+
\end{equation}
and therefore
\begin{eqnarray}\label{c.i}
P_+ \boldsymbol{a} \cdot \dot{\boldsymbol{S}} P_+ & = 
& \frac{p_0}{mc + p_0} c^{-2}
(- (\boldsymbol{a} \cdot \boldsymbol{v}) P_+ 
\dot{\boldsymbol{v}} \cdot \boldsymbol{S}  P_+ +
(\boldsymbol{a} \cdot \dot{\boldsymbol{v}}) P_+ \boldsymbol{v} \cdot \boldsymbol{S} \, P_+)
\nonumber \\ & = & \frac {p_0}{mc + p_0} c^{-2}
P_+ (\boldsymbol{a} \times (\boldsymbol{v} \times \dot{\boldsymbol{v}}))
\cdot \boldsymbol{S} P_+  \\ & = &
- \frac{p_0}{mc + p_0} c^{-2} P_+ \boldsymbol{a} \cdot (\boldsymbol{S} \times
(\boldsymbol{v} \times \dot{\boldsymbol{v}})) P_+
\nonumber \\ & = & - \frac {e}{mc} \frac{1}{1 + \gamma}
P_+ \boldsymbol{a} \cdot (\boldsymbol{S} \times (c^{-1} \boldsymbol{v}
\times \boldsymbol{E} + c^{-2} \boldsymbol{v} \times (\boldsymbol{v} \times
\boldsymbol{B})))P_+\,. \nonumber
\end{eqnarray}
For the no name term we obtain, using again (\ref{c.f}),
\begin{eqnarray}\label{c.j}
\lefteqn{ \hspace{-1cm}- i \frac{e}{2 p_0} P_+ 
[\gamma_5  \gamma_0  \boldsymbol{\gamma}
\cdot \boldsymbol{B}, \boldsymbol{a} \cdot \boldsymbol{S}] P_+ }
\nonumber  \\  &=& \frac{e}{p_0} P_+ \boldsymbol{\gamma} \cdot
(\boldsymbol{B} \times \boldsymbol{g}) \gamma_5 P_+ \nonumber \\
& = & \frac{e}{p_0} (P_+ (\boldsymbol{B} \times \boldsymbol{g}) \cdot 
\boldsymbol{S} P_+ -
\frac{p_0}{mc + p_0} c^{-2} \boldsymbol{v} \cdot (\boldsymbol{B} 
\times \boldsymbol{g})
P_+ \boldsymbol{v} \cdot \boldsymbol{S} P_+)
 \nonumber \\ & = & \frac{e}{mc} P_+ \boldsymbol{a}
\cdot [\frac{1}{\gamma} (\boldsymbol{S} \times \boldsymbol{B}) + \frac{1}{1 + \gamma}
c^{-2}(\boldsymbol{S} \times ( \boldsymbol{v} \times (\boldsymbol{v} \times \boldsymbol{B})))] P_+ \,.
\end{eqnarray}
We add (\ref{c.i}) and (\ref{c.j}) and insert in Eq. (\ref{c.e}). 
Using (\ref{b.n}) we finally obtain
\begin{equation}\label{c.k}
\frac{d}{dt} \boldsymbol{a} \cdot \boldsymbol{s} = \frac{e}{mc} \boldsymbol{a}
\cdot [\boldsymbol{s} \times (\frac{1}{\gamma} \boldsymbol{B} - 
\frac{1}{1 + \gamma} c^{-1}
\boldsymbol{v} \times \boldsymbol{E})] \, ,
\end{equation}
which is the BMT equation for $g = 2$.
\begin{appendix}
\section{Appendix}\label{d}
\setcounter{equation}{0}
We want to understand the solution of the Dirac equation with
slowly varying potentials as in (\ref{a.c}). Changing to the rescaled
space-time variables, denoted again by the same symbols, the Dirac
equation becomes
\begin{equation}\label{d.a}
i \,\varepsilon \hbar  \frac{\partial}{\partial t} \psi = [\gamma_0
\boldsymbol{\gamma} \cdot(- i \varepsilon \hbar  \nabla_x - \frac{e}{c}
\boldsymbol{A} (\boldsymbol{x})) + m c \gamma_0 + e \phi (\boldsymbol{x})] \psi.
\end{equation}
Since only the combination $\varepsilon  \hbar$ appears, we may set
$\hbar = 1$ at the expense that formulas do not look right
dimensionally.

At this point, it is more convenient to abstract from the
particular form of the Dirac equation and to consider a 
general
unitary evolution governed by
\begin{equation}\label{d.b}
i \varepsilon \frac{\partial}{\partial t} \psi = H^\varepsilon \psi
\end{equation}
with $\psi$ an $n$-component spinor, $\psi \in  \bigoplus^n_{j=1} L^2 (\IR^d).$
To define $H^\varepsilon$ we start from a symmetric $n \times n$
matrix-valued function $H (\boldsymbol{q}, \boldsymbol{p})$ on the classical phase space
$\Gamma = \IR^d \times \IR^d.$ Then, in spirit, $H^\varepsilon = H(\boldsymbol{x}, -i \varepsilon
\nabla_x)$. While for the Dirac operator this substitution is
unambiguous, in general it is not, and we define $H^\varepsilon$
through the Weyl quantization
\begin{equation}\label{d.c}
H^\varepsilon \psi (\boldsymbol{x}) = (2 \pi)^{- d}\int d^dy \int d ^d\xi 
 H
\left(\frac {\boldsymbol{x} + \boldsymbol{y}}{2} ,
\varepsilon \boldsymbol{\xi} \right)e^{i(\boldsymbol{x}-\boldsymbol{y})\cdot \boldsymbol{\xi}}
\psi(\boldsymbol{y})\,.
\end{equation}
Next we have to list the properties of the symmetric matrix-valued
Hamiltonian function $H$. We assume that it has the spectral
decomposition
\begin{equation}\label{d.d}
H(\boldsymbol{q},\boldsymbol{p}) 
= \sum^m_{j=1} h_j (\boldsymbol{q},\boldsymbol{p}) 
P_j (\boldsymbol{q}, \boldsymbol{p})\, .
\end{equation}
$h_j$ are the real band energies. We order them as $h_j \le h_{j+1}$
and we assume that bands do not cross which means that
\begin{equation}\label{d.e}
h_j (\boldsymbol{q},\boldsymbol{p}) < h_{j+1} (\boldsymbol{q}, 
\boldsymbol{p})\, ,
\end{equation}
$j=1, \dots , m-1$. If bands cross, then the situation becomes
considerably more complicated. We refer to \cite{hage} for the case
of two nondegenerate bands touching each other at a single point.
$P_j (\boldsymbol{q},\boldsymbol{p})$ are orthogonal, symmetric projections. Since $H$ is
assumed to be smooth in $\boldsymbol{q},\boldsymbol{p}$ and by (\ref{d.e}), their degeneracy must
be independent of $\boldsymbol{q},\boldsymbol{p}$,
\begin{equation}\label{d.f}
{\rm tr} P_j (\boldsymbol{q},\boldsymbol{p}) = d_j
\end{equation}
with degeneracy $d_j$. We have $\sum^m_{j=1} d_j = n$. $d_j = 1$
is a scalar band.
We study (\ref{d.b}) in the limit $\varepsilon \rightarrow 0$ through the
evolution of Wigner density matrices, which for the pure state
$\psi_{t}$ is defined by
\begin{equation}\label{d.g}
W_{t}^\varepsilon (\boldsymbol{q},\boldsymbol{p}) = 
(2 \pi)^{-d} \int d^d \xi \psi_{t}
(\boldsymbol{x} -\frac{\varepsilon}{2} \boldsymbol{\xi})
\psi_{t}^{*}(\boldsymbol{x} + \frac{\varepsilon}{2} \xi)
e^{i \boldsymbol{p} \cdot  \boldsymbol{\xi}}
\end{equation}
and for a general density matrix by the corresponding incoherent
superposition.
$W_{t}^\varepsilon$ is a $n \times n$ matrix-valued function on phase
space. Using the evolution law (\ref{d.b}) for $\psi_{t}$ and expanding the
right hand side of (\ref{d.g}) in $\varepsilon$, one obtains
\begin{equation}\label{d.h}
\frac{\partial}{\partial t} W^\varepsilon _{t} =  - \varepsilon^{-1} i
[H, W_{t}^\varepsilon] + \frac{1}{2} (\{W_{t}^\varepsilon , H\} - 
\{H, W_{t}^\varepsilon \}) +
R_{t}^\varepsilon
\end{equation}
with a remainder $R_{t}^\varepsilon$ which vanishes as 
$\varepsilon \rightarrow 0$
 \cite{gera}. Note that $H$ is the given $n \times n$
matrix-valued hamiltonian function.
In (\ref{d.h}) the first piece of the generator is rapidly oscillating
while the second piece is of order one. Such a situation has been
studied abstractly by Davies \cite {davi}. He proves that, for any $t 
\neq 
0$,
$W^\varepsilon _{t} \rightarrow W_{t}$ in the limit $\varepsilon \rightarrow 0$ 
and that $W _{t}$ commutes with $H$, $[H, W_{t}] = 0$. The order one piece
of the generator is projected onto the invariant subspace of the
evolution generated by $- i [H, \cdot]$. This means that in the
limit $\varepsilon \rightarrow 0$ (\ref{d.h}) goes over to
\begin{equation}\label{d.i}
\frac{\partial}{\partial t} W _{t} = \sum^m_{j=1}  P_j
\frac{1}{2} (\{W _{t}, H \} - \{H, W _{t} \}) P_j
\end{equation}
with initial conditions given through $\lim_{\varepsilon
\rightarrow 0} \sum^m_{j=1} P_j W_{0}^\varepsilon
P_j = W_{0}$.
Note that for all $t$ we have $[P_j, W_{t}] = 0$, $ j = 1, \dots , m.$
It is instructive to rewrite (\ref{d.i}) in a somewhat different form.
We insert the spectral decomposition of $H$,
\begin{eqnarray}\label{d.j}
\lefteqn{\hspace{-0,5cm}\frac{1}{2} \sum^m_{j=1} P_j (\{W, H\}- \{H, W \}) 
P_j}
\\ \hspace{-1cm} &=&\frac{1}{2} \sum^m_{j=1} \sum^m_{\ell=1} P_j
(\{W, h_\ell P_\ell \} - \{h_\ell P_\ell, W\}) P_j
\nonumber \\ \hspace{-1cm} &=& - \sum^m_{j=1} P_j \{h_j,
W\} P_j + \frac{1}{2} \sum^m_{j=1} \sum^m_{\ell=1} h_\ell P_j
(\{W, P_\ell\} - \{P_\ell, W\}) P_j \nonumber \,.
\end{eqnarray}

To continue we need two identities. The first one reads
\begin{equation}\label{d.k}
P_j \{h_j,P_j \} P_j = 0
\end{equation}
which follows from $\{h_j,P_j\} = \{h_j, P^2_j\} = \{h_j, P_j\} P_j + 
P_j \{h_j, P_j\}.$
Therefore, using that $P_j W = WP_j,$
\begin{eqnarray}\label{d.l}
P_j \{h_j, W\} P_j &=& \{h_j, P_j W P_j\} - P_j W \{h_j, P_j\} -
\{h_j, P_j\} W P_j \nonumber \\ &=& \{h_j, P_j W P_j\} - [P_j W
P_j , [P_j, \{h_j, P_j \}]]\,.
\end{eqnarray}

The second identity is derived from
\begin{equation}\label{d.m}
A \{B, C\}-\{A, B\} C = \{A B, C\} - \{A, B C\}\, ,
\end{equation}
compare with \cite{gera}. Then
\begin{eqnarray}\label{d.n}
P_j \{P_j, W\} - \{P_j, P_j\} W &=& \{P_j, W\} - \{P_j, P_j W\}\, ,
\nonumber \\ W \{P_j, P_j\} - \{W, P_j\} P_j &=& \{W P_j, P_j\} -
\{W, P_j\}
\end{eqnarray}
and
\begin{eqnarray}\label{d.o}
P_j( \{W, P_j\}  -  \{P_j , W\} )P_j = W P_j \{P_j, P_j\} P_j
- P_j \{P_j , P_j\} P_j W \nonumber \\ - P_j (P_j \{P_j , W\} -
\{P_j, W\} P_j - \{P_j W, P_j\} + \{P_j, W P_j\}) P_j 
\nonumber \\ \hspace{-3cm}= [W,
P_j \{P_j, P_j\} P_j ]\, ,
\end{eqnarray}
where we use $[W, P_j]= 0$ and again (\ref{d.m}). In addition, again
by (\ref{d.m}),
\begin{equation}\label{d.p}
P_1 \{P_2, P_2\} = - \{P_1, P_2\} (1-P_2)\,,\;\; \{P_2, P_2\} P_1 = - (1 - P_2)
\{P_2, P_1 \}\, .
\end{equation}
Then, by (\ref{d.m}) and $P_1 P_2 = 0$
\begin{eqnarray}\label{d.r}
P_1 \{P_2, W \} - \{P_1, P_2\} W &=& - \{P_1, P_2 W\}\, , \nonumber \\
W  \{P_2, P_1\} - \{W, P_2\} P_1 &=& \{W P_2, P_1\}
\end{eqnarray}
and
\begin{eqnarray}\label{d.s}
P_1 (\{W , P_2\} &-& \{P_2 , W\} ) P_1
\nonumber \\ \quad &=& P_1 (W \{P_2, P_1 \} -
\{P_1, P_2 \} W + \{P_1, P_2 W\} - \{W P_2, P_1 \} P_1 \nonumber
\\ \quad &=& W P_1 (1-P_2) \{P_2, P_1\} P_1 - P_1 \{P_1, P_2 \} (1-P_2)
P_1 W \nonumber \\ \quad &=& - [W, P_1 \{P_2, P_2\} P_1 ]\,,
\end{eqnarray}
where used that $P_1 W = P_1 W $, $ P_1 (1 - P_2) = P_1$, (\ref{d.r}), and
\begin{equation}\label{d.t}
P_1 \{W P_2, P_1\} - \{P_1, W P_2\} P_1 = \{P_1, P_2 W P_1\} -
\{P_1 W P_2, P_1\} = 0,
\end{equation}
which follows from (\ref{d.m}), $P_2 W = W P_2,$ and $P_1 P_2 = 0$.
Let us define
\begin{equation}\label{d.u}
H^{(j)}_s = -i [P_j, \{h_j, P_j\}] - \frac{i}{2} h_j P_j \{P_j,
P_j\} P_j + \frac{i}{2} \sum^m_{\ell=1,  \ell \not= j} h_\ell P_j \{P_\ell,
P_\ell\} P_j\,.
\end{equation}
Then we have transformed (\ref{d.i}) into
\begin{equation}\label{d.v}
\frac{\partial}{\partial t} W_{t} = \sum^m_{j=1} (- \{h_j, P_j W_{t}
P_j\} - i [H^{(j)}_s , P_j W_{t} P_j]).
\end{equation}

If we assume that the initial $W_{0}$ is concentrated in band $j$, 
i.e. $P_j W_{0}P_j
=W_{0}$, and solve
\begin{equation}\label{d.w}
\frac{\partial}{\partial t} W_{t} = - \{h_j , W_{t}\} - i [H^{(j)}_s, W_{t}]\,,
\end{equation}
then $P_j W_{t} P_j = W_{t}$ at all times. If $W_{0}$ is a pure state
for all $\boldsymbol{q}, \boldsymbol{p}$, 
$W_{0} (\boldsymbol{q},
\boldsymbol{p}) = | \psi_{0}(\boldsymbol{q},\boldsymbol{p}) \rangle 
\langle
\psi_{0}(\boldsymbol{q},\boldsymbol{p})|$ 
with $P_j \psi_{0} = \psi_{0}$, then this structure is maintained in 
the course of time, $P_j \psi _{t}
= \psi_{t}$, and $\psi_{t}$ solves
\begin{equation}\label{d.x}
\frac{\partial}{\partial t} \psi_{t} = - \{h_j, \psi_{t}\} - i 
H^{(j)}_s \psi_{t}\,.
\end{equation}
In particular, if initially $W_{0}(\boldsymbol{q},\boldsymbol{p}) =
\delta (\boldsymbol{q} - \boldsymbol{q}^0) \delta (\boldsymbol{p} - 
\boldsymbol{p}^0)| \varphi_{0}\rangle\langle \varphi_{0}|$
with $\varphi_{0} \in P_j \IC^n$, then at any other time
\begin{equation}\label{d.y}
W_{t} (\boldsymbol{q},\boldsymbol{p}) = \delta (\boldsymbol{q} - 
\boldsymbol{q}_{-t})
\delta (\boldsymbol{p} - \boldsymbol{p}_{-t}) |\varphi_t\rangle\langle 
\varphi_t|\, ,
\end{equation}
where $(\boldsymbol{q}_t, \boldsymbol{p}_t)$ is the solution to
\begin{equation}\label{d.z}
\dot{\boldsymbol{q}} = \{h_j, \boldsymbol{q}\} = \nabla_p h_j\,,
\quad \dot{\boldsymbol{p}} = \{h_j, \boldsymbol{p}\} = - \nabla_q h_j
\end{equation}
with initial conditions $\boldsymbol{q}^0, \boldsymbol{p}^0.$ 
Thus the wave packet
propagates along the classical orbit with the band function $h_j$ as
hamiltonian. $\varphi_t$ evolves according to the time-dependent
spin hamiltonian $H^{(j)}_s (t) = H^{(j)}_s(\boldsymbol{q}_t,\boldsymbol{p}_t)$
of the $j$-th band as
\begin{equation}\label{d.aa}
 i \frac{d}{dt} \varphi_t = H^{(j)}_s (t) \varphi_t \, .
\end{equation}
Note that $P_j (\boldsymbol{q}_t, \boldsymbol{p}_t) \varphi_t = \varphi_t.$

The final answer (\ref{d.y}), (\ref{d.aa}) has one rather surprising feature.
We focus on the $j$-th band. To compute the $\boldsymbol{q}, \boldsymbol{p}$ 
evolution we only
have to know the band function $h_j$ of the band under
consideration. The first term in the spinor hamiltonian just
ensures that, as the wave packet propagates, the spinor does not
move out of the $j$-th band subspace. Clearly, $[P_j, \{h_j, P_j\}]$
knows only about $h_j$ and $P_j$, as does the second term of $H^{(j)}_s.$
However, the third term depends on all the other bands. Thus if we
modify $H^\varepsilon$ keeping the $j$-th band fixed, then the spinor
motion, and only it, will be modified. Physically, one might have
been tempted to argue that the other bands are separated from the
band $j$ by a large energy and therefore modifying them should leave
the dynamics in the $j$-th band unaffected. Our computation shows
that the actual behavior is otherwise.
\end{appendix}

\end{document}